\documentclass[dvips,usenatbib]{mn2e}
\usepackage[T1]{fontenc}
\usepackage[latin9]{inputenc}
\usepackage{amsmath}
\usepackage{amsfonts}
\usepackage{amssymb}
\usepackage[dvips]{graphicx}
\usepackage{natbib}
\usepackage{hyperref}            
\defcitealias{sheldonI}{\scshape{S09}b}
\defcitealias{sheldonII}{\scshape{J07}}
\defcitealias{sheldonIII}{\scshape{S09}a}

\newcommand{\der}{\mathrm{d}}

\title[Tracing mass and light in the Universe]{Tracing mass and light in the Universe: where is the dark matter?}
\author[N. A. Bahcall and A. Kulier]{
Neta A. Bahcall and Andrea Kulier\thanks{E-mail: \href{mailto:akulier@princeton.edu}{akulier@princeton.edu}}\\
Department of Astrophysical Sciences, Princeton University, Princeton, NJ 08544}

\begin{document}

\date{\today}

\pagerange{\pageref{firstpage}--\pageref{lastpage}} \pubyear{2013}

\maketitle

\label{firstpage}

\begin{abstract}

How is mass distributed in the Universe? How does it
compare with the distribution of light and stars? We address these questions by examining
the distribution of mass, determined from weak lensing observations, and starlight, around $>10^{5}$ SDSS MaxBCG groups and
clusters as a function of environment and scale, from deep inside
 clusters to large cosmic scales of $22 \,  h^{-1}$ Mpc. The observed cumulative mass-to-light profile,
$M/L \, (< r)$, rises on small scales, reflecting the increasing $M/L$ of the central
bright galaxy of the cluster, then flattens to a nearly constant ratio on scales above $\sim 300  \, h^{-1}$ kpc,
where light follows mass on all scales and in all environments.  A trend of
slightly decreasing $M/L \, (r)$ with scale is shown to be consistent with the varying
stellar population following the morphology-density relation. This suggests that
stars trace mass remarkably well even though they represent only a few percent of the
total mass. We determine the stellar mass fraction and find it to be nearly constant
on all scales above $\sim 300 \,  h^{-1}$ kpc, with $M_{*}/M_{tot} \simeq 1.0\pm0.4\%$. We 
further suggest that most of the
dark matter in the Universe is located in the large halos of individual galaxies 
($\sim 300$ kpc for $L^{*}$ galaxies); we show that the entire $M/L \, (r)$ profile --- from groups and
clusters to large-scale structure --- can be accounted for by the aggregate masses
of the individual galaxies (whose halos may be stripped off
but still remain in the clusters), plus gas. We use the
observed mass-to-light ratio on large scales to determine the mass density of the Universe:
$\Omega_{m} = 0.24 \pm 0.02 \times b_{M/L}^{2}  =  0.26 \pm 0.02.$
\end{abstract}


\section{Introduction}

Understanding the distribution of mass on large scales is a fundamental quest in
cosmology. How does it compare with the distribution of stars,
light, and gas?  We know that individual galaxies are surrounded by large
dark matter halos, and that groups and clusters of galaxies are dominated by
dark matter comprising five to ten times more mass than baryonic matter
(stars and gas).  It is generally believed that the
relative contribution of dark matter increases with scale ---  from galaxies, to
groups, to clusters, and to large-scale structure; larger scale systems are
believed to contain more dark matter, relative to light or stars, than do galaxies
(\citealt{ostriker1974, rubin1978, davis1980}; and more recently
\citealt{guo2010,leauthaud2011}). In this paper we investigate the distribution of mass,
light, and stars and explore how light and stars trace mass
 as a function of environment and scale. We discuss the implications
 for the disribution of dark matter
 and for the stellar mass fraction.

One of the classical methods to investigate the distribution of mass in the Universe
is to compare it directly with the distribution of stellar light:  how does mass follow
light?  This method was first used on large scales by \citet{zwicky}; this was followed by
many additional investigations (see \citealt{bahcall1995, bahcall2000, carlberg1996,
sheldonIII}, and references therein).  The comparison between the
distribution of mass and light is represented by the mass-to-light ratio, $M/L$,
which can be studied as a function of scale and environment.  \citet{ostriker1974},
\citet{davis1980}, and others showed that $M/L$ increases systematically
with scale, from the small scale of galaxies to the larger scale of groups,
clusters, and larger scale structure, indicating a growing dominance of dark matter
with scale (see also \citealt{guo2010}, and references therein). \citet{bahcall1995}
investigated the overall $M/L \, (< r)$ function from galaxies to large scales,
separating galaxies into ellipticals (older) and spirals (younger). They showed that
$M/L \, (< r)$ rises from the small scales of galaxies up to a few hundred kpc,
reflecting the large dark matter halos around galaxies, then flattens to a constant
value, where light approximately traces mass, with
comparable relative contribution of dark matter on all scales. \citet{bahcall1995} 
thus suggest that most of the dark matter in the Universe may be located in the 
large dark matter halos around individual galaxies and that the dark matter in 
groups and clusters may simply be the sum of their individual galaxy members (plus gas). 
More recently, \citet{sheldonIII} used gravitational lensing observations of clusters from the
Sloan Digital Sky Survey (SDSS) and found a similar trend of increasing $M/L$
with radius that flattens on scales of several Mpc. The flattening of $M/L$ on
large scales indicates that light follows mass on these scales and the $M/L$ ratio
approaches a mean cosmic value \citep{bahcall1995, bahcall2000, tinker2005,
sheldonIII}.

In this paper we use the SDSS observations of weak lensing mass and 
the observed distribution of light around 132,473 stacked groups and clusters of galaxies at $0.1 < z < 0.3$
\citep{sheldonI, sheldonII, sheldonIII} to analyze the mass-to-light profile 
as a function of environment, from small groups of a few galaxies to the
richest clusters, and as a function of scale, 
from the small scale of 25 kpc inside clusters
to large cosmic scales of $\sim30 \,  h^{-1}$ Mpc.  These scales reach to
$\sim 20 - 40$ virial radii of the systems, well into the large-scale cosmic environment.
 We investigate how light traces mass on these scales, how the $M/L$ profile is
affected by the varying stellar population,
and how the stars trace the underlying mass distribution.  We estimate the
approximate stellar mass fraction as a function of environment and scale.  We show
that light and stars trace mass on scales above several hundred kpc, and that
the mean stellar mass fraction is nearly constant on all these scales in all environments, 
$M_{*}/M_{tot} \simeq 1.0\pm0.4\%$ (\S \ref{d4}).  
We further show that most of the dark matter may be located in the large
halos around individual galaxies (\S \ref{d3}).

We discuss the data in \S \ref{data} and the analysis and results in \S \ref{results}.
We investigate the contribution of individual galaxies to the total $M/L(r)$ function in
\S \ref{d3}, determine the distribution of the stellar mass fraction in \S \ref{d4}, and calculate
$\Omega_{m}$ in \S \ref{d5}.
We summarize our conclusions in \S \ref{conclusions}. We use a flat
$\Lambda$CDM cosmology with $\Omega_{m} = 0.27$, $\Omega_{\Lambda} = 0.73$, 
and $H_{0} = 100h$ km/s/Mpc (where $h = 0.7$ should be
inserted for $\Lambda$CDM; \citealt{spergel2007}). 

\section{Data}
\label{data}

\subsection{Cluster Sample}
\label{data1}

The MaxBCG cluster catalog \citep{maxbcgI, maxbcgII} was obtained from the Sloan 
Digital Sky Survey (SDSS; \citealt{sdss}) data release 4 \citep{dr4}. 
 The cluster finder is based on the red-sequence method; it maximizes
the likelihood that a galaxy is a brightest
cluster galaxy (BCG) at the center of an overdensity of red-sequence galaxies \citep{maxbcgI}.
All clusters are selected from a 7500 deg$^{2}$ region on the sky and 
are in the photometric redshift range $0.1 < z < 0.3$.  The 
cluster richness $ N_{200}$ is defined by the number of galaxies on the red sequence 
with rest-frame $i$-band luminosity $ L_{i} >  0.4 L^{*} $ located within a radius $r_{200}^{gals}$
from the BCG (where the 
$i$-band $L^{*}$ is the $z=0.1$ value from \citealt{blanton2003b}, corresponding to $M_{*} -  
5\log(h) =  - 20.82 \pm 0.02$, and $r_{200}^{gals}$ 
is the radius within which the local galaxy overdensity is 200; see \citealt{hansen2005}). 
The radius $r_{200}^{gals}$, which is used only to define the cluster richness $N_{200}$, differs from the $r_{200}$ 
we use throughout this paper, which defines a mass overdensity of 200 times the critical density.  
The published catalog 
contains 13,823 clusters with $ N_{200} \ge 10$; \citet{sheldonI} (hereafter \citetalias{sheldonI}) augment the 
catalog to include small groups of galaxies with $ N_{200} \ge 3$, resulting in
a sample of 132,473 groups and clusters with $N_{200} \ge 3$. The 
cluster photometric redshifts are accurate to $0.004$, with a scatter of $\Delta z \sim 
0.01$ for $ N_{200} \ge 10$, degrading to $\Delta z \sim 0.02$ for the poorest systems,
with the same accuracy.

\subsection{Lensing Mass Measurements}
\label{data2}

The cluster sample was partitioned into 12 richness bins 
ranging from small groups with $ N_{200} = 3$ to 
the richest clusters with $ 71 \le N_{200} \le 220$ (\citealt{sheldonII};
 hereafter \citetalias{sheldonII}).
The number of clusters per bin decreases from 58,788 for the
poorest bin to 47 for the richest.
The weak lensing mass 
measurements were carried out by \citetalias{sheldonI} and \citetalias{sheldonII}  on the stacked clusters in 
each richness bin, all centered on the cluster BCG.  Because lensing is not 
sensitive to a uniform mass distribution (``mass sheet''), the 
measured mass reflects the mean mass density of the lens sample above the mean 
density of the Universe.

The average tangential shear of background galaxies due to lensing by the 
foreground stacked clusters was measured by \citetalias{sheldonI} 
and used to calculate
the mass density distribution around the clusters as a function of projected 
radius from the center out to scales of $30 \,  \, h^{-1}$ Mpc. 
\citetalias{sheldonI} made corrections to the density profile for contamination of the lensed 
sample by cluster members and residual additive biases in the tangential shear. The mass density 
contrast of the lensing clusters in each richness bin was measured in logarithmically 
spaced bins of projected clustercentric radius from $25 \,  \, h^{-1}$kpc to $22 \,  \, h^{-1}$Mpc. 
The 2D density contrast profile was then deprojected to a 3D density excess profile
$\Delta \rho (r) \equiv \rho(r) - \bar{\rho}$  using an Abel inversion (see \citetalias{sheldonII} for details). 
This measurement reflects the cluster-mass cross-correlation function
times the mean density of the Universe, $\Delta\rho(r) = \xi_{cm}(r)\bar{\rho}$.
This was used to calculate the cumulative excess mass $\Delta M (<r)$ within each radius $r$,
as well as to find $r_{200}$, the radius within which the average mass density is 200
times the critical density. 

We correct the mass profiles presented by 
\citet{sheldonIII} (hereafter \citetalias{sheldonIII}) 
for an improved photometric redshift distribution of the lensed 
background galaxies; a full description is given in \S \ref{data5}.

\subsection{Luminosity Measurements}
\label{data3}

The light distribution around the stacked clusters, using galaxy luminosities
as defined below, was measured (by \citetalias{sheldonIII}) 
as a function of radius for the same scales as above,
from $25 \,  h^{-1}$ kpc to $30 \,  h^{-1}$ Mpc.
A uniform background was 
subtracted using similar measurements around random points.  The luminosity measured
reflects the luminosity density of the systems above the mean, $ \Delta\ell(r)
 = \ell(r) - \bar{\ell}$, thus representing the cluster-light cross-correlation function as 
a function of radius, $\Delta \ell(r) = \xi_{c\ell}(r)\bar{\ell}$, comparable to the mass measurements described above.  

All galaxy luminosities used are in the $i$-band, with K-correction
applied to bring them to the mean cluster 
redshift of $z = 0.25$; these are denoted $^{0.25}i$. 
All galaxy magnitudes are SDSS model magnitudes.  
A volume and magnitude limited sample 
of galaxies was chosen with $z < 0.3$ and $ L_{^{0.25}{i}} > 10^{9.5} h^{-2} L_{\odot} = 0.19 L^{*}_{^{0.25}i}$
\citepalias{sheldonIII}.

The projected luminosity profiles discussed above were 
inverted to obtain the 3D luminosity density profile \citepalias{sheldonIII}, similar to the mass 
inversion. This was then integrated to obtain the excess cumulative luminosity $\Delta L (<r)$ within the same
clustercentric radial bins as above, from $25 \, h^{-1}$kpc to $22 \,  h^{-1}$Mpc. Because the 
luminosity density was not measured within the innermost $25 \,  h^{-1}$kpc where the 
BCG is located, the average luminosity of the BCG in each stacked richness bin 
was added to the cumulative luminosity, so that the central $L_{BCG}$ is 
included in the light (see \citetalias{sheldonIII} for more details). 

\subsection{Mass to Light Ratio}
\label{data4}

The excess mass within radius $r$ from the center of the stacked clusters, 
$\Delta M(<r)$, divided by the excess luminosity within radius $r$, $ \Delta L(<r) $,
gives the mass-to-light ratio
\begin{equation}
\frac{\Delta M}{\Delta L}(<r) = 
\frac{\int_{0}^{r}\der r \, [\rho(r) - \bar{\rho}] \, r^{2}}{\int_{0}^{r} \der r \, [\ell(r) - \bar{\ell}] \, r^{2}} = 
\frac{\int_{0}^{r}\der r \, \bar{\rho} \,\xi_{cm} \,r^{2}}{\int_{0}^{r}\der r \, \bar{\ell} \,\xi_{c\ell} \,r^{2}}.
\end{equation}
On small scales of virialized systems, this represents the mean cluster mass-to-light 
ratio.  On large scales, as the density approaches the mean, this measures
approximately the mean mass-to-light ratio of the Universe and reflects how mass traces light on large scales.  
A small correction factor 
reflecting the bias of the galaxy light tracers relative to the mass relates the 
observed large-scale asymptotic $ \Delta M(<r)/\Delta L(<r) $ to the mean cosmic value $<M/L>$:
\begin{flalign}
\left(\frac{\Delta M}{\Delta L}\right)_{asym} =& \left\langle\frac{M}{L}\right\rangle b_{M/L}^{-2} \notag \\
b_{M/L}^{-2} =& \frac{b_{cm}}{b_{c\ell}}\frac{1}{b_{\ell m}^{2}}.
\end{flalign}
The bias factor $b_{M/L}$ depends on the ratio of the bias of clusters relative to mass and light,
$ b_{cm}/b_{c\ell}$, which is near unity since the cluster bias essentially cancels
out in this ratio. The bias of the galaxy tracers relative to the mass, 
$b_{\ell m}$, is $ \approx 1$ for galaxies near $L^{*}$ \citep{sheth1999,seljak2004,zehavi2011}
and varies very slowly for galaxies below $L^{*}$ \citep{tegmark2004,zehavi2011}.
Since the luminosity threshold of 
the galaxy tracers is only $ 0.19L^{*} $ (and the mean luminosity 
of all the galaxies within $10 \, h^{-1}$Mpc is $ 0.65 L^{*}$), this implies a bias of 1.05
(for $\sigma_{8} = 0.83$; \citealt{zehavi2011}). 
Thus, on large scales, $\Delta M (<r)/\Delta L (<r) $ should approach a constant 
value, representing the cosmic mass-to-light ratio (with only minor bias correction). 

\subsection{Corrections and Uncertainties}
\label{data5}

We correct the lensing masses of \citetalias{sheldonII} and \citetalias{sheldonIII} using the correction 
from \citet{rozo2009} that reflects an improved treatment of the 
photometric redshift distribution of the background source galaxies
based on a detailed analysis by \citet{mandelbaum2008}. This
correction increases the lensing mass by 
$18\% \pm 2\%$ (\textit{stat.}) $\pm 2\%$ (\textit{sys.}).
We also correct the $r_{200}$ values by the small correction needed 
($\sim 5\%$) due to this mass increase.

The mass and luminosity determinations discussed above use BCGs as the 
center for the stacked clusters. If the BCG is slightly offset from the 
center of the cluster (e.g., \citealt{niederste2010}), both the lensing mass and luminosity
in the central regions will be slightly underestimated
(relative to the central mass and light of the cluster dark matter halo)
 although the effect will partially
cancel in the mass-to-light ratio. The effect is strongest 
near the center of the clusters, as well as in poor groups for which
miscentering is more likely to occur. However, the effect is negligible on large
scales. Discussion of the effect can be found in \citetalias{sheldonIII},
\citet{mandelbaum2008} and \citet{tinker2012}.
We do not correct for this effect 
as the corrections are uncertain and are typically small (except in the poorest groups);
 they are not important for our main goal 
of understanding the behavior of the mass-to-light profile on larger scales.
Our results thus reflect the mass-to-light ratio around BCG galaxies, 
not necessarily around the center of the dark matter halos, especially for small 
groups.

We further note that unless otherwise stated (e.g., \S \ref{d2} - \ref{d4}) the luminosity accounts 
only for galaxies above the luminosity threshold of $L_{^{0.25}i} > 10^{9.5} h^{-2} L_{\odot}$,
as discussed above. It does not 
include the entire luminosity of the systems. This is discussed in \S \ref{d2}.

\section{Analysis and Results}
\label{results}

We combine the twelve stacked bins of groups and clusters (\S \ref{data})
into three broader richness bins to study the overall  $ M/L $ as a function of radius around poor 
($3 \le N_{200} \le 8$), intermediate ($9 \le N_{200} \le 25$), 
and rich ($26 \le N_{200} \le 220$) systems. We use
the same radial bins as in \citetalias{sheldonIII}. We investigate both 
the cumulative $\Delta M (< r)/\Delta L (< r)$, and the 
local, differential mass-to-light profile, $\Delta m(r)/\Delta \ell (r)$.
For simplicity, we refer to these as $M/L \, (<r)$ and $m/\ell \, (r)$,
respectively.

\begin{figure}
  \begin{center}
    \includegraphics[width=\columnwidth]{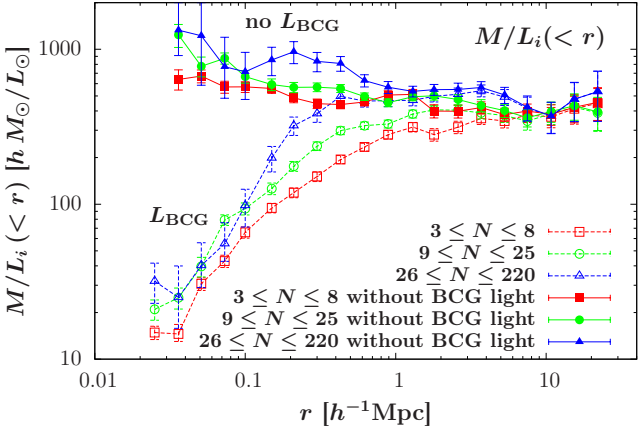}
    \caption{The cumulative mass-to-light ratio, $M/L_{i}\, (<r)$, for poor 
	($3 \le N_{200} \le 8$), intermediate ($9\le N_{200} \le 25$), and rich
	($26 \le N_{200} \le 220$) clusters. The dashed curves and related points show $M/L$
	including both the light and mass of the central BCG, whereas the solid curves and related points
	show $M/L(< r)$ excluding the BCG light. }
    \label{graph:graph100}
  \end{center}
\end{figure}

The integrated mass-to-light profile $M/L \, (<r)$
 is presented in Figure \ref{graph:graph100} as a function
 of radius from $25  \, h^{-1}$kpc to $22 \, h^{-1}$Mpc for each of 
the three richness bins. The observed increase of $M/L \, (<r)$ with radius on small 
scales is caused by the central BCG galaxy which dominates the cluster
luminosity in the central regions (typically $\lesssim 300$ kpc):   
these regions reflect primarily $M/L \, (<r)$ of the BCG galaxy.
The three top curves in Figure \ref{graph:graph100} present
$M/L \, (<r)$ when the mean BCG luminosity is excluded from
each richness bin, revealing the impact 
of the BCG. A comparison of the two sets of curves shows that 
the BCG luminosity accounts for the increasing $M/L \, (<r)$
in the central regions, and has negligible effect on the mass-to-light ratio on 
large scales. When the BCG light is excluded, the integrated mass-to-light profile is 
nearly flat on all scales, except for the rise on small scales that 
results from excluding the light (i.e., $ L_{BCG}$) but not the mass. 
If we further exclude the mass within the central region (see below), $ M/L \, (<r) $ 
 flattens on all scales. On large 
scales $M/L \, (<r)$ is essentially independent of cluster richness;  all environments,  from the 
smallest groups to the richest clusters,  exhibit the same overall mass-to-light ratio, 
reaching a universal value on scales of a few Mpc. 
 On intermediate scales of $\sim 0.5 - 5 \, h^{-1}$Mpc,
 while $M/L \, (<r)$ is nearly flat, 
small differences are observed as a function of richness and radius. We discuss 
these trends below and show that they are consistent with the varying 
mean stellar population age as a function of richness and radius.  

The rise in $ M/L \, (<r)$ on small scales is considerably faster for rich
 clusters than for poor groups (Figure \ref{graph:graph100}). 
 Rich clusters reach a nearly flat 
$M/L \, (<r)$ distribution at only $\sim 0.3  \, h^{-1}$Mpc  (typically near the edge 
of the bright BCG).  Poor groups have a slow rise of $M/L \, (<r)$ with radius: the BCG 
luminosity dominates the group's luminosity and thus $ M/L \, (<r) $ to
scales of a few Mpc.  This is because the BCG luminosity, while 
known to increase with cluster richness ($L_{BCG} \sim M_{200}^{0.30} \sim N_{200}^{0.38}$;
\citealt{hansen2009}),
is considerably more dominant in poor groups ($L_{BCG}/L_{200} \sim
 M_{200}^{-0.53} \sim N_{200}^{-0.67}$; \citealt{hansen2009}).  This 
is easily understood:  in a poor group of a few galaxies, the BCG typically 
contains most of the group luminosity; as the group's luminosity $\Delta L(<r)$ is integrated over radius, the BCG light 
dominates the group luminosity to large scales.  In rich 
clusters, on the other hand, the BCG is brighter but constitutes only a small fraction of 
the total cluster luminosity at nearly all scales outside the innermost region.  

\begin{figure}
  \begin{center}
    \includegraphics[width=\columnwidth]{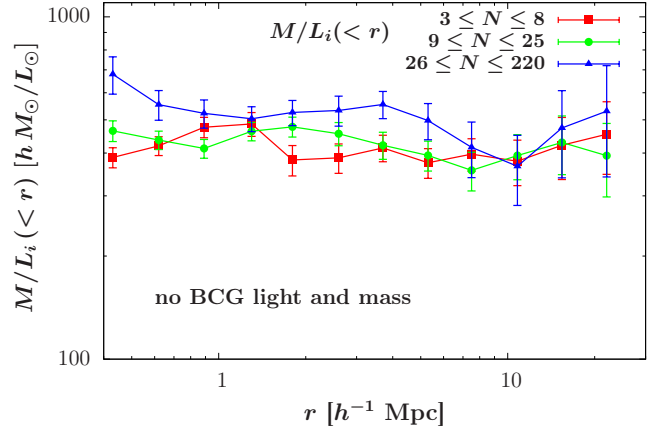}
    \caption{The cumulative mass-to-light ratio, $M/L_{i}\, (<r)$, outside
the central regions, with the BCG light and central mass excluded (see \S \ref{results}).}
    \label{graph:graph200}
  \end{center}
\end{figure}

Because the effect of the BCG is dominant in the innermost regions, and our interest 
is mostly in understanding the general behavior 
of $ M/L \, (<r) $ on larger scales, we present in Figure \ref{graph:graph200}
 the cumulative 
$M/L (<r)$ profile excluding both the central BCG
luminosity (as in Fig. \ref{graph:graph100}) as well as the central mass.
For the latter, we exclude all the mass within a central radius of 50 $ \, h^{-1}$ 
kpc for the poorest systems (which have the smallest BCGs) up to 150 $ \, h^{-1}$ 
kpc for the richest clusters.  Selecting somewhat different radii for the 
central mass exclusion has only a small effect on the overall $ M/L \, (<r) $ 
profiles (e.g., compare with Figure \ref{graph:graph100} where no central mass has been excluded); 
it does not affect the main results discussed below.   
The $ M/L \, (<r) $ profile outside the innermost regions (Figure \ref{graph:graph200}) shows a nearly 
flat distribution on all scales, 
for all systems rich and poor, from the surprisingly small scale of few 
hundred kpc to the largest cosmic scales at 22 $ \, h^{-1}$ Mpc. The fact that the 
distribution is nearly flat on all scales indicates that stars, which make up
only a few percent of the total mass, trace the distribution of mass well.

While $ M/L \, (<r) $ is nearly flat, we observe a small 
trend with richness and with radius:  $ M/L \, (<r) $ increases slightly with richness 
at a given radius (on scales of $\lesssim 5$ Mpc),  and decreases slightly with radius 
(up to few Mpc) for all richnesses. As we show below, these trends are 
consistent with the different mix of stellar populations --- old in E/S0
galaxies and younger in spiral galaxies --- as a function of radius and richness.
 Following the density-morphology relation 
\citep{dressler1980,dressler1997,postman1984,vanderwel2008,bamford2009}, the fraction of E/S0 galaxies is
 high in high-density regions (rich clusters), and decreases with radius 
and richness to the lower density regions on larger scales and in poorer groups.
Since early-type galaxies are dominated by 
an old stellar population with negligible recent star formation, their $ M/L_{i} $ ratio 
is larger (by a factor of $\sim 2$; see below) than that of the younger stellar population of spiral galaxies, whose 
luminosities are dominated by bright young stars.  As the fraction of spirals 
increases with radius, the integrated $ M/L $ of the cluster 
decreases slowly with radius, as observed.  The 
small increase of $ M/L $ from poor to rich clusters at a given radius (Figure \ref{graph:graph200})
reflects partly the same effect: at a given radius a more massive cluster
is generally at a smaller multiple of its virial radius, and thus a 
higher density, so it is dominated
 by a larger fraction of
old E/S0s with higher $M/L$. 
  
\begin{figure}
  \begin{center}
    \includegraphics[width=\columnwidth]{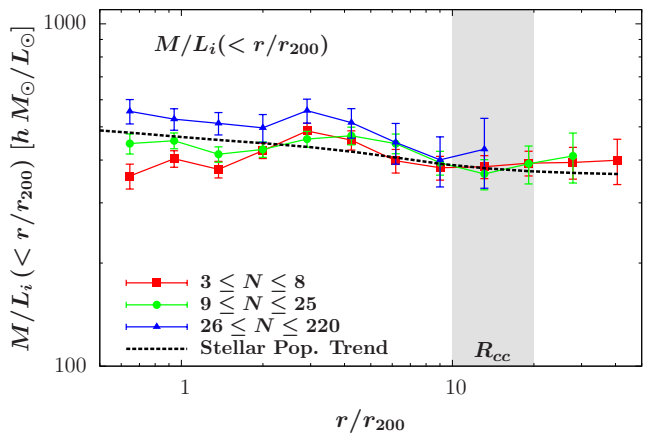}
    \caption{The cumulative mass-to-light ratio, $M/L_{i}\, (<r)$, 
	outside the innermost cluster regions (BCG mass and light excluded).
	Same as Figure \ref{graph:graph200}, but plotted against the clustercentric radius in units of
	$r_{200}$. The dashed line shows the
	expected trend of the varying stellar population age
	as a function of scale (\S \ref{results}). The vertical band shows 
	the location of the group and cluster correlation scale.}
    \label{graph:graph300}
  \end{center}
\end{figure}

\begin{figure}
  \begin{center}
    \includegraphics[width=\columnwidth]{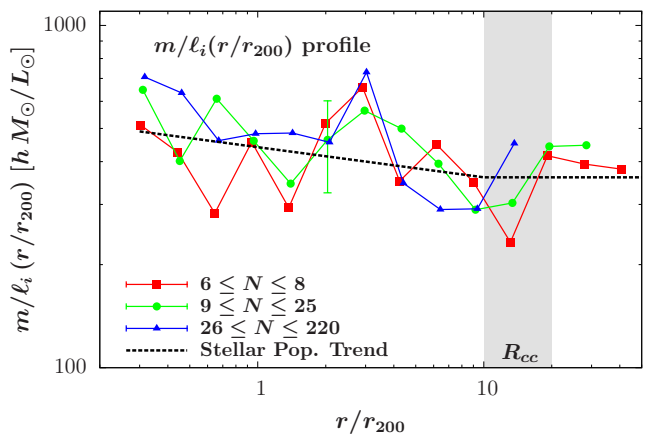}
    \caption{The local mass-to-light profile, $m/\ell_{i}\, (r)$,
	plotted against the clustercentric radius in units of $r_{200}$. 
	 The dashed line shows the expected trend of the varying stellar population (see text).
	For clarity, only a representative error bar is shown.}
    \label{graph:graph400}
  \end{center}
\end{figure}

Figures \ref{graph:graph300} and \ref{graph:graph400} present the cumulative and local (differential)
mass-to-light profiles, respectively, as a function of radius in units of the physical 
scale of the system, $r/r_{200}$,
from small scales to nearly 40$r_{200}$.
In Figure \ref{graph:graph300}, we again exclude the BCG luminosity and the innermost mass in order to 
better understand the overall trend of $ M/L \, (<r) $  without the dominant 
impact of the BCG. The differential analysis 
(Figure \ref{graph:graph400}) is independent of the BCG,
except in the very central region, since it reflects the local $ m/\ell \, (r) $ ratio 
at any radius.
 The results in both figures highlight the similarity of 
$M/L \, (r/r_{200})$ for all systems and on all scales.
Mass and light appear to follow each other --- especially when accounting
for the stellar population age --- from deep inside
clusters (outside their BCG cores)
out to large cosmic scales.

\begin{figure*}
  \begin{center}
    \includegraphics[width=\textwidth]{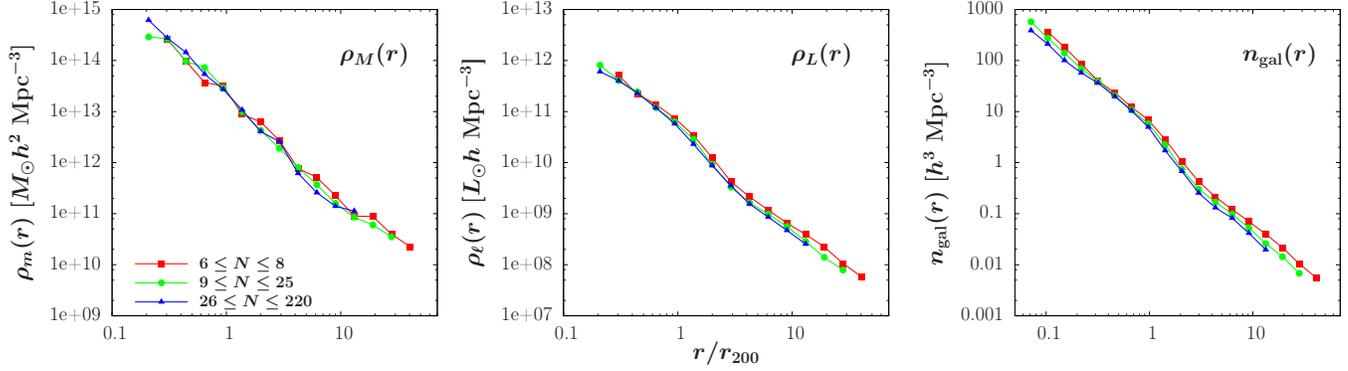}
    \caption{The density profiles of mass, $\rho_{m}(r)$, luminosity $\rho_{\ell}(r)$, and galaxy number density $n_{\mathrm{gal}}(r)$
	as a function of $r/r_{200}$ for clusters of different richnesses. (The $n_{gal}(r)$ plot has a
	slightly different horizontal axis scaling.)}
    \label{graph:graph450}
  \end{center}
\end{figure*}

Figure \ref{graph:graph450} presents the observed profiles of the mass density, luminosity density, 
and galaxy number density for all environments as a function of $r/r_{200}$. The profiles are nearly identical, 
independent of richness. This similarity of profiles is equivalent to the similar $ m/\ell \, (r) $ profile
observed for all systems as shown above. 
Since $r/r_{200}$ corresponds to a similar overdensity for 
all systems, no significant stellar population effect is expected between the different
richness bins, and none is observed.

The richest clusters show a 
slightly higher cumulative $ M/L \, (<r) $ at $r/r_{200} \lesssim 4$ (Figure \ref{graph:graph300}).
This is caused by a somewhat higher $M/L$ at the very center
of the clusters, which may reflect 
a higher central mass 
concentration relative to luminosity in the innermost regions of the richest 
systems (e.g., more massive or extended BCGs than our central mass exclusion and/or loss of luminosity due to 
merging).

The $ m/\ell \, (r) $ profile (Figure \ref{graph:graph400})
is similar for all richness groups and on all scales starting from deep inside clusters ($r \sim 0.3r_{200}$); 
the profile decreases slowly with radius, by a factor of $\sim1.5$, from small scales up to several $r_{200}$ and
remains flat thereafter to nearly $40r_{200}$. 
This trend is consistent with the stellar population mix,
which we quantify below. 

We use the mean observed relative stellar $M_{*}/L_{i}$ ratio of early-type versus
spiral galaxies of $ (M_{*}/L_{i})_{E}/(M_{*}/L_{i})_{S} \simeq 2$ 
\citep{kauffmann2003, bell2003, blanton2007, yi2008, gallazzi2011, leauthaud2011} .
This is consistent with \citet{bc2003} single stellar population
synthesis models with ages 10 Gyr versus 4 Gyr, the approximate
mean ages of stellar populations in E/S0 and spiral galaxies
(albeit with much scatter; see \citealt{trager2000,proctor2002,thomas2005,kuntschner2010,zhu2010,roediger2011}).
We combine this ratio with the mean fraction of early
and late type galaxies using the density-morphology relation discussed above. Here we use a typical E/S0 fraction of 
40\% on large-scales ($> 10r_{200}$), increasing to 90\% in the central high-density regions
of groups and clusters ($<0.4r_{200}$). The resulting mass-to-light ratio then decreases by a factor 
of $\sim1.4$ from the centers of clusters to the lower densities on large scales.   
This is presented by the dashed lines in Figures \ref{graph:graph300}
and \ref{graph:graph400} (for the cumulative and differential functions, 
respectively). This expected trend agrees 
well with the observed $ M/L $ profile for all systems, 
suggesting that the small decline in $ M/L $ as a function of scale
can be accounted for by the different stellar population.  Clusters have a strong spatial 
auto-correlation function and are located in high-density regions with a higher 
density of older galaxies that extends to large scales; this is likely the 
reason that the high $ m/\ell \, (r) $ ratio persist to several $r_{200}$
before dropping to a lower more constant field value on large scales.  The typical cluster 
correlation scale is  $\sim10-20 r_{200}$ \citep{bahcall2003} for both 
poor and rich clusters; this is shown by the vertical band in Figures \ref{graph:graph300}
and \ref{graph:graph400}. While 
$M/L$ reaches the cosmic value on these large scales, we emphasize that when 
the stellar population age is accounted for, all systems exhibit the same mean 
mass-to-light ratio on all scales. Light, or more precisely stellar mass, 
thus traces the total mass remarkably well on nearly all scales.
This is discussed further in \S \ref{d4}. 

\begin{figure}
  \begin{center}
    \includegraphics[width=\columnwidth]{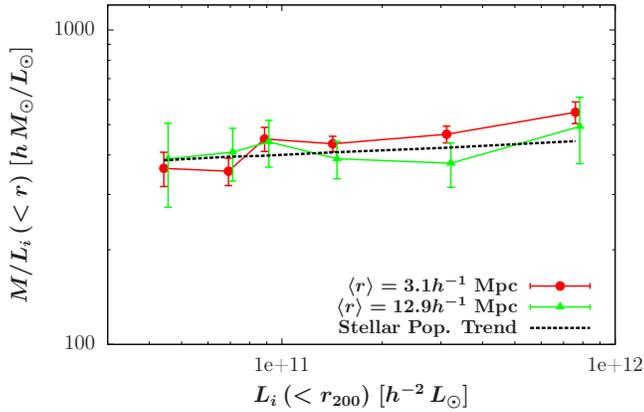}
    \caption{The cumulative mass-to-light ratio, $M/L_{i}\, (<r)$,
	as a function of the mean cluster luminosity within $r_{200}$,
	$L_{i} (<r_{200})$ (a proxy for cluster richness or mass).
	The mass-to-light ratio is shown within two radii, $3.1  \, h^{-1}$ Mpc
	and $12.9  \, h^{-1}$ Mpc. The dashed line represents the expected trend of 
	the varying stellar age (for $\sim 3.1 \, h^{-1}$ Mpc; see text). The BCG mass
	and light are excluded.} 
    \label{graph:graph500}
  \end{center}
\end{figure}

\begin{figure}
  \begin{center}
    \includegraphics[width=\columnwidth]{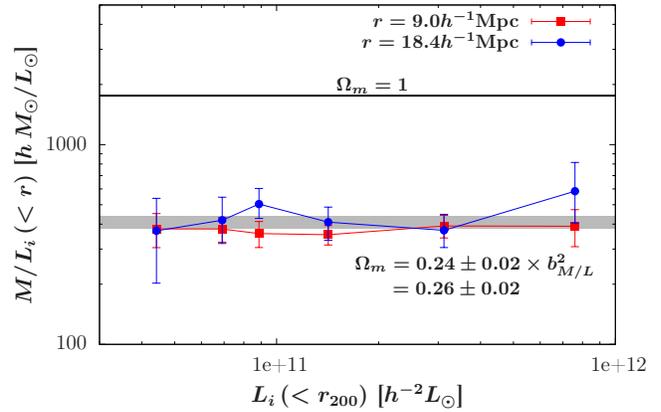}
    \caption{$M/L_{i} (< r)$, including the BCG mass and light, as a function
	of richness at large scales ($9.0  \, h^{-1}$Mpc and $18.4  \, h^{-1}$Mpc). The horizontal axis is
	$L_{i} (< r_{200})$, the cumulative light within $r_{200}$,
	a proxy for the cluster richness or mass. $M/L$ is constant
	on these scales, independent of the central environment. The gray band shows the mean 
	$\left\langle M/L_{i} \right\rangle = 409 \pm 29 \, h \, M_{\odot}/L_{\odot}$. 
	This corresponds to a cosmic mass-density of $\Omega_{m} = 0.24 \pm 0.02 b_{M/L}^{2}
	= 0.26 \pm 0.02$ (see \S \ref{d5}).}
    \label{graph:graph600}
  \end{center}
\end{figure}

An alternate way to view the behavior of $ M/L \, (<r) $ for different environments and 
scales is presented in Figure \ref{graph:graph500}.  Here we show $ M/L \, (<r) $ as a function of 
mean cluster luminosity within $r_{200}$ (a proxy for cluster richness or mass), for several 
radii; this is an inverse of the previous plots of $ M/L \, (<r) $ presented 
for a few richness groups.  
Figure \ref{graph:graph500} shows the direct dependence of $ M/L \, (<r) $ on the richness (mass) of the 
central environment.
The cumulative $M/L \, (<r)$ is presented within two radii,
$3.1  \, h^{-1}$ Mpc and $12.8  \, h^{-1}$ Mpc. The mass and light of the central BCG 
have been excluded in Figure \ref{graph:graph500} as they
were in Figures \ref{graph:graph200} and \ref{graph:graph300}, 
although the effect of the BCG is negligible on the larger scale ($\sim 12.9$ Mpc).
Each point in the cluster luminosity $L_{i}(<r_{200})$ is an average of two
of the twelve original richness bins. The stellar 
population effect as a function of $L_{i}(<r_{200})$ for the 3.1 Mpc radius is shown by the
dotted line; it agrees well with the observed $M/L$ function, except for the 
richest clusters, which as described above contain an increased amount of
mass compared to light at their very centers. Here, the
slow increase of $M/L$ with richness due to the stellar population trend
 is caused by the fact that the same physical radius
corresponds to a different galaxy overdensity in environments of different richness.

Figure \ref{graph:graph600} presents the integrated $ M/L \, (<r) $ versus luminosity 
(richness)  for all systems within the large scales of 9 and 18.4 $ \, h^{-1}$ Mpc. 
The $M/L$ contains all the mass and light including that of the BCG. The 
integrated $ M/L \, (<r) $ ratio on these scales is constant, independent of the 
central environment; the mass-to-light ratio reaches a mean cosmic value of $409\pm23 h M_{\odot}/L_{\odot}$.  As seen in 
the previous figures, the same constant value is reached even within the inner 
parts of clusters when the stellar population mix is accounted for.

\subsection{Discussion of The Mass to Light Function}
\label{d1}

All the measurements discussed above are centered on the BCG, which 
may be slightly offset from the center of the dark matter 
halo of the cluster. Our results therefore reflect the distribution of $ M/L $ 
around BCGs out to large cosmic scales and should be so interpreted. Furthermore, 
for scales outside the innermost regions of clusters,
the possible mis-centering has negligible impact on $M/L$.
The poorest groups have the highest uncertainty in terms of cluster
centering, luminosity uncertainties \citepalias{sheldonIII}, and
 mass uncertainties because
of the weaker lensing signal. Therefore, the 
poorest groups ($N \lesssim 9$) are prone to larger uncertainties.

Previous results found that $M/L(<r_{200})$ increases with cluster richness 
(mass),  approximately as $M/L(<r_{200}) \sim  M_{200}^{0.2-0.3}$ 
 (e.g., \citealt{girardi2000, bahcall2002, lin2004, popesso2007, sheldonIII}).  Here we show 
that this is mostly the effect of the BCG luminosity:  the $r_{200}$ scale of the 
poorest groups is nearly the same size as the BCG galaxy (a few hundred kpc), 
therefore the group $M/L(<r_{200})$ is dominated by the lower $ M/L \, (<r) $ of the BCG, while
the BCG luminosity of rich clusters is negligible at the cluster scale of 
$r_{200} \sim 1$ Mpc. Using the observed relations ($L_{BCG} \propto M_{200}^{0.3}$ and
 $ L_{BCG}/L_{200} \sim M_{200}^{-0.53}$;  
\S \ref{results}; \citealt{hansen2009}) we find $M/L(<r_{200}) =  M_{200}/L_{200} \sim M_{200}^{0.2}$,
  as is indeed observed.  Comparing the $ M/L $ profile outside the innermost region  
(at $r/r_{200} \gtrsim 0.3$), we show that $ M/L(r/r_{200})$ is nearly independent of 
richness (Figures \ref{graph:graph300} and \ref{graph:graph400}). The 
nearly flat distribution of $M/L(r)$ persists on all scales, from $r \sim 0.3r_{200}$ ($\sim
80$ kpc in small groups and $\sim 300$ kpc in large clusters) out to 
nearly $30  \, h^{-1}$ Mpc. The small decrease in the mass-to-light ratio
as a function of $r/r_{200}$ is consistent with a decreasing stellar
population age as a function of local galaxy density (\S \ref{results}).
This effect, combined with the fact that the physical radius $r$ corresponds
to a different galaxy density in clusters of different richnesses,
also accounts for the observed trend in $M/L \, (<r)$ 
as a function of cluster richness at fixed $r$ (Figure \ref{graph:graph500}).
This agreement implies that the underlying mass and light distribution, when 
accounting for the stellar population age, follow each other remarkably well 
on all scales. It further suggests that the stellar mass fraction, $ M_{*}/M $, 
is nearly constant on these scales (see \S \ref{d4}). 

\subsection{Total Luminosity}
\label{d2}

The luminosities discussed above do not represent the total luminosities since
they include only galaxies above the threshold
of  $ L_{i}^{0.25} = 10^{9.5} h^{-2} L_{\odot} = 0.19L^{*}_{^{0.25}i}$ 
(\S \ref{data}). The luminosity contributed by fainter galaxies is not 
included, and neither is the diffuse 
intracluster light (ICL) --- integrated light from individual stars in the cluster
potential \citep{zibetti2005}. While this does not affect the self-consistent
results discussed above, the total luminosity
should be accounted for when comparing our results with other measurements.
We estimate these additional luminosity contributions below.

Galaxies below the threshold of $ 10^{9.5} h^{-2} L_{\odot}  = 0.19 \, L^{*} $ 
are estimated to contribute an additional $36\%$
to the current luminosity, using the observed Schechter luminosity
function of SDSS galaxies which has a faint-end slope of $-1.21$ in the $i$-band
(\citetalias{sheldonIII}; consistent with \citealt{montero2009} and
\citealt{blanton2001}). 
The radial distribution of faint galaxies is assumed to be similar to that of the other
galaxies as suggested by the SDSS observations of \citet{hansen2009} showing that
the galaxy luminosity function in groups and clusters is independent of radial
scale.

The diffuse intracluster light (ICL)
from individual stars, thought to be stripped from
galaxies as a result of gravitational interactions within groups and clusters,
has been measured in detail for SDSS clusters at $z \approx 0.25$.
\citet{zibetti2005} used stacked images of 683 clusters to measure the ICL
in the $i$-band out to 700 kpc. They find that the ICL contributes $\sim 11\%$
of the luminosity in groups and clusters,
nearly independent of cluster richness and BCG luminosity.
The ICL is more concentrated than the galaxy distribution. \citet{tal2011} measure 
the ICL around 42,000 stacked SDSS Luminous Red Galaxies (LRGs) that are located in the centers 
of groups and clusters. They find an ICL contribution of $\sim 20\%$.  
We adopt an ICL correction of $15\%$ additional luminosity (i.e., $\sim 13\%$ 
of the total cluster luminosity). 

We add the above luminosity of faint galaxies 
($36\%$) and the intracluster light
($15\%$); they decrease the $M/L$ ratios by these 
factors (1.36 and 1.15) on the group and cluster scales (i.e., $<r_{200}$), and by $36\%$
(a factor of 1.36) everywhere
else. This corrected total $m/\ell (r)$ profile is presented in Figure \ref{graph:graph800}.
It enables us to compare the results with other measurements
and with the contribution from individual galaxies, as discussed in \S \ref{d3}.

\section{Mass to Light Contribution from Individual Galaxies}
\label{d3}

It is generally believed that groups and clusters have significantly more 
dark matter (relative to light) than individual $ \sim L^{*} $ galaxies
(e.g., \citealt{ostriker1974,davis1980,guo2010}). \citet{bahcall1995} and \citet{bahcall1998} suggested, however,
 that clusters and groups do not contain significantly more dark matter per unit 
light than do galaxies, but instead most of the dark matter 
resides in the large halos of individual galaxies; these make up the dark
matter observed in groups, clusters, and large-scale structure.
We investigate this further below:  how much 
of the observed mass, and thus $M/L$, is contributed by individual galaxies
(plus gas). We use the mean
observed mass-to-light ratio of typical isolated $ \sim L^{*} $ elliptical and spiral galaxies, 
combined with the density-morphology relation (\S \ref{results}) to estimate 
the amount of mass and  $ M/L $ contributed by individual galaxies 
to the observed $ M/L $ profile on all scales and in all environments.

As discussed in \S \ref{results},  the $M/L(< r)$ function flattens to a constant value
 at approximately $\sim 300 \, h^{-1}$ kpc.  This scale is comparable to the virial radius 
of bright $L^{*}$ galaxies ($\sim 250-300 \, h^{-1}$ kpc at the virial overdensity of 
$\sim95\rho_{c}$ relevant for $\Lambda$CDM).  If galaxy halos extend to these scales, 
as suggested by observations (below), could the dark matter of individual galaxies 
(including these large halos), plus the known  gas component, account for all or most
 of the mass observed in the $M/L_{i}(< r)$ function, from small scales inside clusters
 to the large cosmic scales of nearly $30  \, h^{-1}$ Mpc?  To test this we use, for simplicity, 
a $\sim 300 \, h^{-1}$ kpc halo radius around $L^{*}$ galaxies, with $M/L_{i}$ values that are
 consistent with observations:  $M/L_{i} (\lesssim 300 \, h^{-1}) \sim 150 h M_{\odot}/L_{\odot}$
 for spirals, and, with a factor of two lower $L_{i}$  for older galaxies (\S \ref{results}), 
an $M/L_{i} (\lesssim 300 \, h^{-1}) \sim 300h$ for early-type $L^{*}$ galaxies (E/S0).  

These values are motivated by observations of individual $L^{*}$ galaxies.  For example, the Milky Way (MW) and M31, both nearly
 $L^{*}$ spiral galaxies, have been measured in detail to dermine their extended masses (see below).
Their luminosities are  
$L_{i}(MW) \sim 3\times10^{10} L_{\odot}$ and $L_{i}(M31) \sim 3.5 \times 10^{10} L_{\odot}$
(\citealt{cox1999} observe $L_{B}(MW) = 2.4\times 10^{10} L_{\odot}$;
\citealt{courteau1999} infer $M_{V}(MW) = -20.9$ and $M_{V}(M31) = -21.2$;
 \citealt{flynn2006} report $L_{I}(MW)$ in the range  $3 - 4 \times 10^{10} L_{\odot}$;
\citealt{tamm2012} find a total M31 luminosity of $3.5 \times 10^{10} L_{\odot}$).
 This luminosity is consistent with the SDSS 
$L^{*}_{i}$ used above ($3.3\times10^{10} L_{\odot}$ when converted to $h = 0.7$). 

The extended mass of the 
Milky Way and M31 have been measured recently using proper motions observed with 
the Hubble Space Telescope (HST) combined with the Timing Argument (TA);
   \citet{vandermarel2012} determine the sum of the two virial masses to be $4.93 \pm 1.6 \times 10^{12} M_{\odot}$, 
consistent with the \citet{liwhite2008} revised TA method that gives $5.27 \pm 0.5 \times 10^{12} M_{\odot}$.
  \citet{boylan2013} combined the Leo-I proper motion with the TA method and 
other measurements to obtain $M(MW)  \sim 1 - 2.4 \times 10^{12} M_{\odot}$,
and \citet{liwhite2008} obtain an estimated virial mass of $M(MW) \sim 2.4 \times 10^{10} M_{\odot}$. 
More recently, \citet{phelps2013} applied the Least Action Method \citep{peebles}
 to the local group galaxies to derive masses that are consistent with the high end of the above values:
  they find $M(MW) = 2.5 \pm 1.5 \times 10^{12} M_{\odot}$ and $M(M31) = 3.5 \pm 1 \times 10^{12} M_{\odot}$; 
their best value for the MW increases to $M(MW) = 3.5 \pm 1.5 \times 10^{12} M_{\odot}$
when data of four external groups is included.  
Considering the above masses and luminosities, while uncertain, we use as approximate values
$M/L_{i} \sim 100(h = 0.7) \approx 150h M_{\odot}/L_{\odot}$ for $\sim L*$ spiral galaxies
 (within radius $\sim 250-300$ kpc).  
   Early-type galaxies, with their fainter $L_{i}$ luminosities (by a factor of about 2), 
thus imply $M/L_{i} (\lesssim 300 \, h^{-1}$ kpc) $\sim 200(h = 0.7) M_{\odot}/L_{\odot} \approx 300h M_{\odot}/L_{\odot}$ 
 for $\sim L^{*}$ E/S0 galaxies.
 This is consistent with recent observations of isolated early-type galaxies using weak and strong gravitational lensing.
  \citet{lagattuta2010} use HST data  to study strong and weak lensing by isolated elliptical galaxies and find a mean 
$M/L_{V} (\lesssim 300 \, h^{-1}$ kpc) = $300 \pm 90 h$, consistent with the value above. 
 \citet{brimioulle2013} conduct a detailed weak-lensing analysis of galaxies from the CFHT Legacy Survey, finding a mean 
$M/L_{r} \sim 287^{+95}_{-85}h$ for red $L^{*}$ galaxies in low-density regions 
(using a truncated isothermal sphere model with an observed best-fit truncation radius of 
$245^{+64}_{-52} \, h^{-1}$ kpc).  They also find a mean $M/L_{r} \sim 178 \pm 22 h$ for all the red and blue galaxies combined 
(with a best-fit truncation radius of $184  \, h^{-1}$ kpc).  

We adopt the above $M/L_{i}(\mathrm{E/S0}) \sim 300h M_{\odot}/L_{\odot}$
 and $M/L_{i}(\mathrm{Sp}) \sim150h M_{\odot}/L_{\odot}$
  within $\sim300$ kpc of individual $\sim L^{*}$ 
galaxies as approximate values to illustrate the contribution of galaxies to the 
dark matter and $M/L(< r)$
 function on all scales.
  To determine how much galaxies with these $M/L_{i}$ ratio contribute
we combine these galaxy $M/L_{i}$ values with the density-morphology 
relation that describes the mean observed fraction of spiral and E/S0 galaxies as a function of density (\S \ref{results}).
 The E/S0 fraction decreases from nearly $\sim90-100\%$ at the high-density regions of clusters to $\sim40\%$ 
in the low-density field on large scales.  This yields an estimate of the mean $M/L_{i}$ contributed by the galaxies. 
 (We note that while the large dark matter halos of individual galaxies may be stripped off in the dense regions of clusters,
and halos may overlap,
 the total mass remains within the cluster potential.) 
 The contribution from galaxies therefore ranges from $M/L_{i} \sim 270-300h$ in the dense regions of clusters,
 decreasing to $\sim 210h$ on large scale (where $\sim40\%$ of galaxies are E/S0s). 
We add the additional gas component that exists on all scales; for this, we use a gas 
component that is $\sim 15\%$ of the total mass, since the cosmic baryon fraction
is $17\%$ \citep{spergel2007}, and the stellar fraction is $\sim 1-2\%$ (\S \ref{d4}).
This is consistent with the extended gas distribution observed in groups and clusters within
their virial radii \citep{rasheed2010}.
Using a lower gas fraction, as observed in the central parts of small groups $(<r_{500})$, or a slightly
varying gas fraction, only affects the results on the smallest scales ($r/r_{200} <
0.7$) of small groups by $\lesssim 10\%$, well consistent with the observed $M/L(r)$ function.
On larger scales, the intergalactic medium (IGM) is expected
to contain a similar gas fraction (\citealt{cen2006}; see also \citealt{prochaska2013}
for the extended gas component observed in the circumgalactic medium around galaxies).

\begin{figure}
  \begin{center}
    \includegraphics[width=\columnwidth]{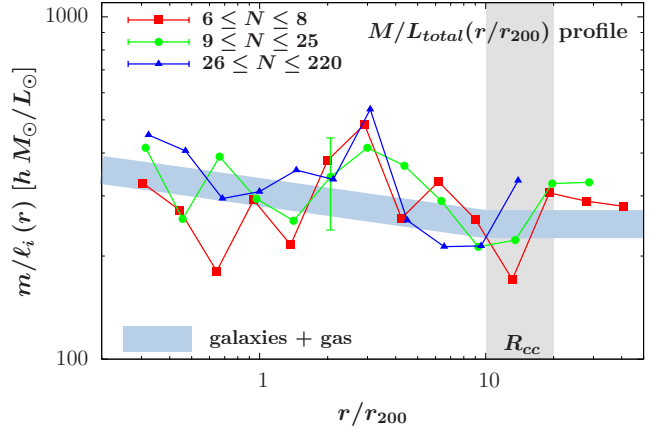}
    \caption{The mass-to-light profile, where $L_{total}$ is the total luminosity
	(including light from faint galaxies below our luminosity threshold plus the diffuse ICL; see \S \ref{d2}). 
	The declining blue band represents the $M/L$ of individual galaxies plus gas (\S \ref{d3}).
	 The galaxies plus gas appear to contribute all or most of the total observed $M/L(r)$ function on all scales
	and environments.}
    \label{graph:graph800}
  \end{center}
\end{figure}

 This sum of galaxy and gas mass represents the mass contributed by individual galaxies (including their dark matter halos)
 plus the intracluster/intergalactic/circumgalactic gas. This contribution ranges from $M/L_{i} \sim340h$ 
inside rich clusters, decreasing to $\sim 250h$ on large scales.  This $M/L_{i}$ 
of individual galaxies plus gas is plotted as the declining
blue band in Figure \ref{graph:graph800}. It agrees remarkably well with the
entire observed  $m/\ell_{i}(r/r_{200})$ profile ---
  from the small scales inside groups and clusters to the largest cosmic scales of nearly $30 \, h^{-1}$ Mpc. 

The comparison in Figure \ref{graph:graph800} shows that individual galaxies and their large 
dark matter halos, plus the intracluster/intergalactic gas, can account for all or most of the 
mass in groups, clusters, and on large scales, reproducing both the total amount of dark 
matter and the overall distribution of mass and light on all scales.  This suggests that 
most of the dark matter in the Universe may be located in the large 
halos of individual galaxies; groups, clusters, and large-scale structure do not appear to contain 
significantly more dark matter (relative to light) than do their $L^{*}$ galaxy 
members. This is consistent with the earlier results of \citet{bahcall1995}.

The values used above for individual galaxies are approximate, with large uncertainties. Using 
lower $M/L_{i}$ values for galaxies will reduce their contribution, and vice versa. 
Faint galaxies, which are known to have higher $M/L$ ratios, will further increase 
the dark matter contribution from individual galaxies.

\begin{figure}
  \begin{center}
    \includegraphics[width=\columnwidth]{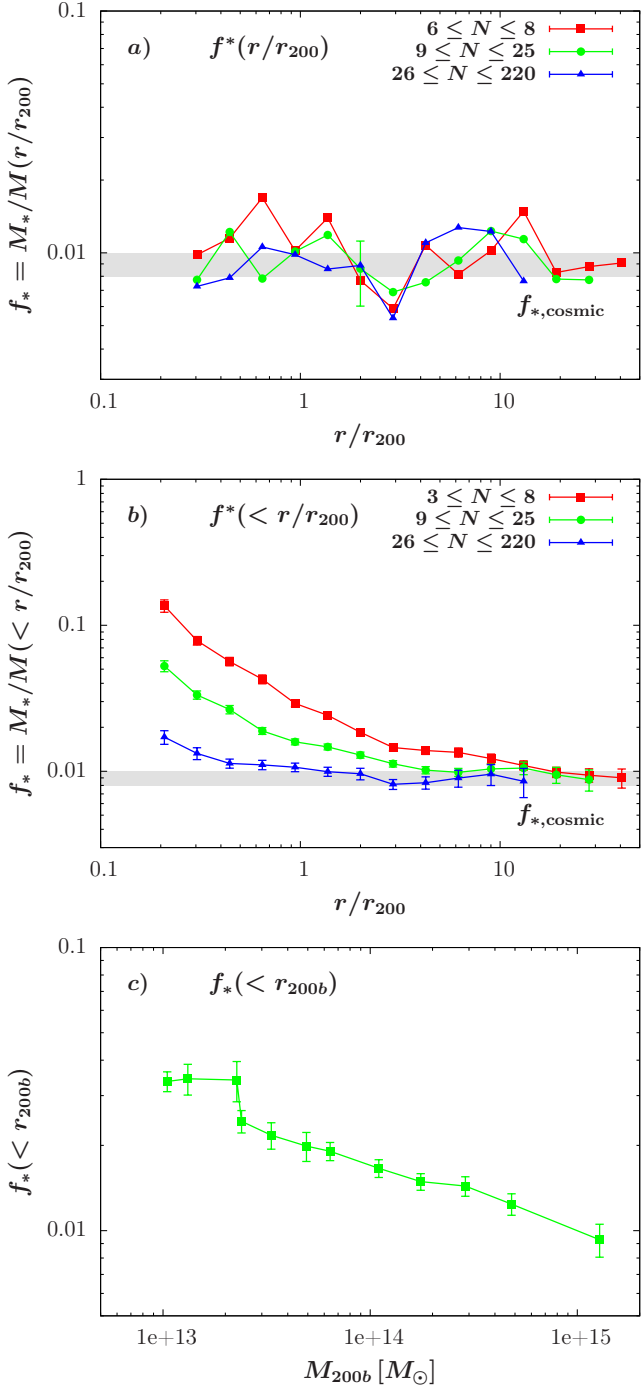}
    \caption{\textit{a}) The local stellar mass fraction $f_{*} (r) = M_{*}/M (r)$
	as function of $r/r_{200}$, for the three
	richness bins. The horizontal band shows the cosmic stellar mass fraction (\S \ref{d4}).
	\textit{b}) As in panel \textit{a}, but for the cumulative stellar mass fraction $f_{*}(<r)$. 
	\textit{c}) The stellar fraction $f_{*}$ within $r_{200b}$ (the radius within which the mean
	density is 200 the matter density) plotted as a function of $M_{200b}$ (the mass
	within $r_{200b}$).}
    \label{graph:graph900}
  \end{center}
\end{figure}

\section{Stellar Mass Fraction}
\label{d4}

The nearly flat distribution of $M/L_{i}$ with radius, especially when accounting for the
age of the stellar population, suggests that the stellar mass fraction may
be similar in all environments, thus causing light to trace mass.  We estimate 
below the implied stellar mass fraction as a function of scale and environment.
We divide the mean observed stellar mass-to-light ratio $M_{*}/L_{i}$ of 
early and late-type galaxies, coupled with their relevant fraction as given
by the density-morphology relation (\S \ref{results}), by our $M/L_{i}$,
including the BCG. This yields the approximate stellar mass fraction $f_{*} = M_{*}/M$:
\begin{flalign}
\label{stellarfraction}
f_{*} (r/r_{200}) = &  M_{*}/M (r/r_{200}) \\ \notag
\simeq &  \left[\frac{M_{*}/L_{i}}{M/L_{i}} (\mathrm{E/S0, S})\right] (r/r_{200})  
\end{flalign}

Estimates of the $M_{*}/L_{i}$ ratios of E/S0s and spirals vary
significantly throughout the literature, 
although most stellar and dynamical mass-to-light estimates
yield $(M_{*}/L_{i})_{E}/(M_{*}/L_{i})_{S} \simeq 2$
 \citep{kauffmann2003, bell2003, blanton2007, yi2008, gallazzi2011, leauthaud2011}.
Stellar mass-to-light ratio determinations can differ as a result of the
precise stellar population synthesis model and data used
to obtain them. \citet{graves2010}
compare several $M_{*}/L_{V}$ estimates for early-type galaxies from different sources,
all corrected to $z = 0$ and a Chabrier initial mass function (IMF) for consistency. A 
similar comparison for $M/L_{I}$ (G. Graves, private communication) reveals that
models based on fits to observational spectra (\citealt{kauffmann2003}, using SDSS data; \citealt{gallazzi2005})
give $M_{*}/L_{I}$(E/S0)$\simeq 2.5 M_{\odot}/L_{\odot}$ (and 
$M_{*}/L_{I}$(Sp)$\simeq 1.3 M_{\odot}/L_{\odot}$), 
as does a single burst star formation model. Models based on fits
to photometric SEDs \citep{bell2003, blanton2007} find lower
values of $M_{*}/L_{I}$(E/S0)$\simeq 2 M_{\odot}/L_{\odot}$ ($M_{*}/L_{I}$(Sp)$\simeq 1 M_{\odot}/L_{\odot}$).
Similarly, \citet{leauthaud2011}, using the COSMOS data,
find $M_{*}/L_{i}$(E/S0)$\simeq 1.7 M_{\odot}/L_{\odot}$ and 
$M_{*}/L_{i}$(Sp)$\simeq 0.9 M_{\odot}/L_{\odot}$. All the 
above use the Chabrier IMF (or are scaled to it).
Adopting the Salpeter IMF increases these mass-to-light ratios
by a factor of $\sim 2$.

Measurements of $M_{*}/L_{i}$ for the Milky
Way and M31, ``typical'' $L^{*}$ spiral
galaxies, find $M_{*}/L_{I}(MW) = 1.3 - 1.6 M_{\odot}/L_{\odot}$ \citep{flynn2006}
or $M_{*}/L_{I}(MW) = 1.4-2.0 M_{\odot}/L_{\odot}$ \citep{mcmillan2011,flynn2006},
and \citet{tamm2012} find $M_{*}/L_{i}(M31) \simeq 2.9 M_{\odot}/L_{\odot}$ for all components
of the galaxy combined. The value for M31 is significantly higher than 
that found for the Milky Way and other spirals above; this may be 
caused by M31's relatively red color for a spiral \citep{tempel2011}.
The $M_{*}/L_{i}$ values for the Milky Way and M31 seem to indicate a higher
$M_{*}/L_{i}$ for spirals than found by some of the models above; as a result,
we adopt the higher range of values from the sources described, using 
$M_{*}/L_{i}$(E/S0)$=2.5 M_{\odot}/L_{\odot}$ and 
$M_{*}/L_{i}$(Sp)$=1.3 M_{\odot}/L_{\odot}$. The results for the stellar fraction discussed below
are proportional to the assumed value of $M_{*}/L_{i}$; 
using different values, the resulting 
stellar mass density and stellar mass fraction will 
scale proportionally. 

Since $M_{*}/L_{i}$ is independent of the Hubble parameter $h$, we 
convert the $M/L_{i}$ ratios presented in the previous section (which were presented in units
of $h$) to the $\Lambda$CDM $h = 0.7$ (thus reducing all $M/L_{i}$ by 0.7).
The total observed mass-to-light ratio --- including the BCG and accounting
for the total luminosity of the system (\S \ref{d2}) --- is then
used to determine $f_{*}$ as described above.
In the top panel of Figure \ref{graph:graph900} we present
the local stellar fraction $f_{*}$ as a function of radius in units
of $r/r_{200}$. The stellar mass fraction is essentially constant
with radius from 0.2 to $40r_{200}$, with a value $f_{*} \simeq 1.0\pm0.4\%$
for all richnesses. Also shown is a horizontal band representing
the average stellar mass fraction of the Universe. For this,
we use the measured luminosity density of $1.61 \pm 0.05\times 10^{8} h L_{\odot} \mathrm{Mpc}^{-3}$
(\S \ref{d5}), increased by a factor of 1.36 to account for 
the light from galaxies below our luminosity threshold (\S \ref{d2}).
We combine this with the $M_{*}/L_{i}$ for spirals 
and E/S0s as discussed above to obtain a cosmic
mean stellar mass density of $\rho_{*} = 2.8 \pm0.2 \times 10^{8} M_{\odot} \mathrm{Mpc}^{-3}$ (for $h=0.7$),
implying $f_{*,\mathrm{cosmic}} = 0.9 \pm 0.1 \%$. This is consistent with 
the mean stellar mass density of $\rho_{*} \simeq 3.1 \pm 0.5 \times 10^{8} M_{\odot} \mathrm{Mpc}^{-3}$ 
($f_{*} = 0.9\%\pm0.2\%$ for $\Omega_{m} = 0.26 \pm 0.02$; \S \ref{d5})
found by \citet{muzzin2013} using the data of \citet{cole2001,bell2003,baldry2012}.

The middle panel of Figure \ref{graph:graph900} shows the
cumulative stellar fraction $f_{*}$ within $r/r_{200}$.
The strong influence of the central BCG is 
clearly seen in the poor groups, as discussed in \S \ref{results},
causing the increase in stellar fraction near the cluster center. At small $r/r_{200}$, 
especially in poor groups, the stellar fraction increases
as the BCG becomes dominant.
On large scales,
systems of all richnesses reach the same cumulative stellar fraction of $f_{*} \simeq 1\%$.
In the lower panel of Figure \ref{graph:graph900}, we present
the stellar fraction within $r_{200b}$, the radius at which the interior mean density
is 200 times the matter density of the Universe, versus $M_{200b}$, the mass within $r_{200b}$.
The increasing stellar fraction at lower mass (smaller $r_{200b}$) reflects
the dominance of the central BCG on this scale (\S \ref{results}). 

\section{ $\Omega_{m}$}
\label{d5}

The observed mass-to-light ratio on large scales can be used to determine the 
mass density of the Universe,  $ \Omega_{m}$  (e.g., \citealt{bahcall1995, bahcall1998}, \citetalias{sheldonIII} 
and references therein).  Figure \ref{graph:graph600} presents the observed $M/L_{i}$  within two 
large scales (9 and 18.4 $  \, h^{-1}$ Mpc)  around all systems, from the poorest groups 
to the richest clusters (\S \ref{results}). The observed $M/L_{i}$  within these large scales is independent 
of the central environment, reaching a representative cosmic value 
(see also \citetalias{sheldonIII}). 
This mean value, shown by the horizontal band in Figure \ref{graph:graph600}, is $409 \pm 29 h M_{\odot}/L_{\odot}$.
Combined with the luminosity density of the Universe in the $ i^{0.25}$ band for galaxies above $0.19L^{*}$,
 $ 1.61\pm0.05\times10^{8} h L_{\odot}/$Mpc$^{3}$ (comoving) \citepalias{sheldonIII}, we find
$\Omega_{m} = 0.24 \pm 0.02 b_{M/L}^{2}$, independent of $h$.  The small 
bias factor, $b_{M/L}^{2}$, depends primarily on the bias of the galaxy tracers relative 
to the underlying mass distribution (see \S \ref{data}).  The galaxies 
used have a low threshold luminosity of $ 10^{9.5} h^{-2} L_{\odot} =  0.19L^{*}$.  The bias of such 
sub-$L^{*}$ threshold galaxies is $b \simeq 1.05$ (for $\sigma_{8} = 0.83$;
increasing to 1.09 for $\sigma_{8} = 0.8$; \citealt{zehavi2011}).   
We thus find
\begin{equation}
\Omega_{m} = 0.24 \pm 0.02 \times b_{M/L}^{2} = 0.26 \pm 0.02.
\end{equation}

\section{Summary and Conclusions}
\label{conclusions}

We compare the distribution of mass and light around 132,473 BCG-centered SDSS
groups and clusters as a function of scale, from small scales inside clusters to
large cosmic scales of nearly $30 \, h^{-1}$ Mpc, and for different richness environments. The
masses are determined from stacked weak gravitational lensing observations and are
used to derive the mass-to-light profile, $M/L_{i}(<r)$; this profile indicates how light
traces mass on all scales and in all environments.  We summarize our main
conclusions below.

1. The $M/L_{i} (<r)$ profile rises with radius on small scales inside groups and
clusters, reflecting the increasing mass-to-light ratio of the central bright BCG
galaxy.  $M/L_{i} (<r)$ then flattens to a nearly constant value, showing that light
follows mass on large scales. This flattening to a nearly constant $M/L_{i}$ ratio begins
at relatively small scales of only a few hundred kpc inside clusters and remains
nearly constant to the largest cosmic scales of $\sim30 \, h^{-1}$ Mpc; this is especially so
when accounting for the varying stellar population age as a function of local
density (see $\#3$ below).  This indicates that light follows mass on all scales (above
$\sim300$ kpc) and in all environments, even inside clusters.  The rise to
the flat constant value in the cumulative $M/L(< r)$ function is considerably slower
in poor groups  because they are dominated by light from the central BCG
galaxy (see $\#2$ below).  

2. The luminosity of the BCG galaxy increases with cluster richness, but its
dominance relative to the total cluster luminosity decreases with richness. 
 This decreasing trend of  $L_{\mathrm{BCG}}/L_{\mathrm{cluster}}$  with cluster richness affects the
cumulative $M/L_{i}$ profile in the central regions of groups and clusters, causing the
slower increase of $M/L_{i} (<r)$ with radius for poor groups, but reaching a constant
$M/L_{i}$ ratio on very small scales of $\sim300$ kpc in rich clusters.   The decreasing
dominance of the BCG with richness suggests that the growth of the central BCG is
less efficient than the growth of the cluster as a whole.  Clusters, which evolve by
merging and accretion of poorer systems, grow faster than do their central BCG
galaxies (also by merging and accretion). 

3.  The small trends observed in the $M/L_{i} (<r)$ profile outside the central
BCG regions --- a slowly decreasing $M/L_{i}$
with radius and richness --- are shown to be consistent with the varying stellar
population as a function of density, following the density-morphology relation.
This stellar population trend results in a slowly decreasing $M/L_{i}$ with radius due to the
increasing population of young spiral galaxies, which have lower $M/L_{i}$
than ellipticals, before flattening to a constant cosmic
value on larger scales.  This indicates that stars, which account for
only a few percent of the total mass, trace  the total mass remarkably well. The
dark matter in the Universe thus follows light, and especially stellar mass, on all scales
above a few hundred kpc.

4. We determine the stellar mass fraction as a function of environment
and scale, $f_{*} = M_{*}/M (r/r_{200})$. We find that the stellar mass fraction is
nearly constant on all scales and all
environments above few hundred kpc, with $f_{*} \simeq 1.0\pm0.4\%$. 
This fraction is consistent with the cosmic stellar mass fraction. 
The fact that stars follow mass so well is the main reason why light 
traces mass on all these scales. 

5. We show that most of the dark matter in the Universe may be located in large dark
matter halos around individual galaxies ($\sim300$ kpc for $L^{*}$ galaxies).  The mass and
mass-to-light ratio of groups, clusters, and large scale structure is consistent
with being contributed by the mass of individual galaxies, including their large
dark matter halos (which may be stripped off inside the dense regions of clusters),
plus the additional intergalactic/intracluster gas. The mass from individual
galaxies, plus gas, appear to be consistent with the entire observed $M/L$
profile on all scales and in all environments (Fig. \ref{graph:graph800}). This suggests that most of the
dark matter in the Universe may be located in the large halos of individual
galaxies;  groups, clusters, and large scale structure are simply made-up by this
dark matter; they do not contain significantly more dark matter (relative to light)
than do the individual galaxies. 

6. The constant $M/L_{i}$ ratio on large scales
represents the universal mass-to-light ratio. This corresponds to a mass-density
parameter of $\Omega_{m} = 0.24 \pm 0.02 \times b_{M/L}^{2}  =  0.26 \pm 0.02$  (where the small galaxy
bias factor is $\sim 1.05$). 

\section*{Acknowledgments}
We thank the referee for the helpful comments on the paper.
A. K. acknowledges support from an NSF graduate research fellowship.

\bibliographystyle{mn2e}

\makeatletter
\let\jnl@style=\rmfamily 
\def\ref@jnl#1{{\jnl@style#1}}%
\newcommand\aj{\ref@jnl{AJ}}
\newcommand\araa{\ref@jnl{ARA\&A}}
\newcommand\apj{\ref@jnl{ApJ}}
\newcommand\apjl{\ref@jnl{ApJ}}
\newcommand\apjs{\ref@jnl{ApJS}}
\newcommand\ao{\ref@jnl{Appl.~Opt.}}
\newcommand\apss{\ref@jnl{Ap\&SS}}
\newcommand\aap{\ref@jnl{A\&A}}
\newcommand\aapr{\ref@jnl{A\&A~Rev.}}
\newcommand\aaps{\ref@jnl{A\&AS}}
\newcommand\azh{\ref@jnl{AZh}}
\newcommand\baas{\ref@jnl{BAAS}}
\newcommand\icarus{\ref@jnl{Icarus}}
\newcommand\jrasc{\ref@jnl{JRASC}}
\newcommand\memras{\ref@jnl{MmRAS}}
\newcommand\mnras{\ref@jnl{MNRAS}}
\newcommand\pra{\ref@jnl{Phys.~Rev.~A}}
\newcommand\prb{\ref@jnl{Phys.~Rev.~B}}
\newcommand\prc{\ref@jnl{Phys.~Rev.~C}}
\newcommand\prd{\ref@jnl{Phys.~Rev.~D}}
\newcommand\pre{\ref@jnl{Phys.~Rev.~E}}
\newcommand\prl{\ref@jnl{Phys.~Rev.~Lett.}}
\newcommand\pasp{\ref@jnl{PASP}}
\newcommand\pasj{\ref@jnl{PASJ}}
\newcommand\qjras{\ref@jnl{QJRAS}}
\newcommand\skytel{\ref@jnl{S\&T}}
\newcommand\solphys{\ref@jnl{Sol.~Phys.}}
\newcommand\sovast{\ref@jnl{Soviet~Ast.}}
\newcommand\ssr{\ref@jnl{Space~Sci.~Rev.}}
\newcommand\zap{\ref@jnl{ZAp}}
\newcommand\nat{\ref@jnl{Nature}}
\newcommand\iaucirc{\ref@jnl{IAU~Circ.}}
\newcommand\aplett{\ref@jnl{Astrophys.~Lett.}}
\newcommand\apspr{\ref@jnl{Astrophys.~Space~Phys.~Res.}}
\newcommand\bain{\ref@jnl{Bull.~Astron.~Inst.~Netherlands}}
\newcommand\fcp{\ref@jnl{Fund.~Cosmic~Phys.}}
\newcommand\gca{\ref@jnl{Geochim.~Cosmochim.~Acta}}
\newcommand\grl{\ref@jnl{Geophys.~Res.~Lett.}}
\newcommand\jcp{\ref@jnl{J.~Chem.~Phys.}}
\newcommand\jgr{\ref@jnl{J.~Geophys.~Res.}}
\newcommand\jqsrt{\ref@jnl{J.~Quant.~Spec.~Radiat.~Transf.}}
\newcommand\memsai{\ref@jnl{Mem.~Soc.~Astron.~Italiana}}
\newcommand\nphysa{\ref@jnl{Nucl.~Phys.~A}}
\newcommand\physrep{\ref@jnl{Phys.~Rep.}}
\newcommand\physscr{\ref@jnl{Phys.~Scr}}
\newcommand\planss{\ref@jnl{Planet.~Space~Sci.}}
\newcommand\procspie{\ref@jnl{Proc.~SPIE}}
\let\astap=\aap 
\let\apjlett=\apjl 
\let\apjsupp=\apjs 
\let\applopt=\ao 
\makeatother

\label{lastpage}

\end{document}